\documentclass[12pt]{iopart}
\usepackage{iopams}

\expandafter\let\csname equation*\endcsname\relax

\expandafter\let\csname endequation*\endcsname\relax

\usepackage{amsmath}

\usepackage{graphicx}
\usepackage{bm}

\usepackage[colorlinks,allcolors=blue]{hyperref}
\begin{document}

\title[Free Electron Laser Generation of X-Ray Poincar\'e Beams]{Free Electron Laser Generation of X-Ray Poincar\'e Beams}

\author{Jenny Morgan$^{1,2}$, Erik Hemsing$^3$,
Brian W. J. M$^{\rm c}$Neil$^{1,2,4}$ and Alison Yao$^1$}
\address{$^1$ University of Strathclyde (SUPA), Glasgow G4 0NG, United Kingdom}
\address{$^2$ Cockcroft Institute, Warrington, WA4 4AD, UK}
\address{$^3$ SLAC National Accelerator Laboratory, Menlo Park, California 94025, USA}
\address{$^4$ ASTeC, STFC Daresbury Laboratory, Warrington, WA4 4AD, UK}

\ead{jenny.morgan@strath.ac.uk}
\vspace{10pt}
\begin{indented}
\item[]\today
\end{indented}

\begin{abstract}

An optics-free method is proposed to generate X-ray radiation with spatially variant states of polarization via an afterburner extension to a Free Electron Laser (FEL).  Control of the polarization in the transverse plane is obtained through the overlap of different coherent transverse light distributions radiated from a bunched electron beam in two consecutive orthogonally polarised undulators. Different transverse profiles are obtained by emitting at a higher harmonic in one or both of the undulators. This method enables the generation of beams structured in their intensity, phase, and polarization - so-called Poincar\'e beams - at high powers with tunable wavelengths. Simulations are used to demonstrate the generation of two different classes of light with spatially inhomogeneous polarization - cylindrical vector beams and full Poincar\'e beams.
\end{abstract}

%
%
%
%
%

\section{Introduction}

Polarization is important when considering light's interaction with matter. The majority of past research has concentrated on light with polarization which does not vary with transverse spatial coordinate, such as linear, elliptical or circular polarization. However, there has been growing interest in vector, or Fully Structured Light (FSL) beams with spatially-varying polarization states~\cite{Zhan:09,Beckley:10,Galvez:12} which can have additional, beneficial properties for a number of applications. For example, beams with radially orientated linear polarization can be focused more tightly than those with spatially homogeneous linear polarization, with applications in laser machining, optical nano-probing, and nano-lithography~\cite{PhysRevLett.91.233901, Rubinsztein_Dunlop_2016,Zhan:09}. Beams with transversely structured polarization have also been shown to propagate more stably in self-focussing nonlinear media~\cite{PhysRevLett.117.233903}.  In general, the ability to control both the intensity and the polarization of FSL beams may provide a useful method for applications in material processing~\cite{Nesterov_2000}, microscopy~\cite{Hao_2010}, and in atomic state preparation, manipulation and detection~\cite{PhysRevLett.85.4482, PhysRevLett.86.5251}.
In this paper, a relatively simple method to generate tunable FSL into the X-ray using Free Electron Lasers (FEL)~\cite{McNeilNatPho2010} is described, opening up new, unexplored areas of atomic and molecular science. One such area is in the field of mirror image chiral molecules, either left or right handed, also called enantiomers. When subjected to FSL, a discriminatory optical force in opposite directions can result for each enantiomer~\cite{chiralReview}.

In the simplest form, FSL beams can be described by a vector superposition of two orthogonally polarised spatial eigenmodes: 
\begin{equation} \label{eq:vector superposition of transverse modes}
\boldsymbol{E}(\boldsymbol r,\boldsymbol\phi)=E_1(\boldsymbol  r,\boldsymbol\phi)\boldsymbol{\hat{e}_1}+e^{i\beta}E_2(\boldsymbol r,\boldsymbol\phi)\boldsymbol{\hat{e}_2}
\end{equation}
where $\beta$ is the phase between the two modes and $\boldsymbol{\hat{e}_1}$ and $\boldsymbol{\hat{e}_2}$ are orthogonal polarization vectors. For cylindrically symmetric beams, a Laguerre-Gaussian ($LG$) set of spatial eigenmodes and circular polarization basis is adopted, where $\boldsymbol{\hat{e}_1}=\boldsymbol{\hat{e}_L}=(\boldsymbol{\hat{e}}_x+i\boldsymbol{\hat{e}}_y)/\sqrt{2}$ and $\boldsymbol{\hat{e}_2}=\boldsymbol{\hat{e}_R}=(\boldsymbol{\hat{e}}_x-i\boldsymbol{\hat{e}}_y)/\sqrt{2}$
correspond to left and right-handed circular polarization vectors, respectively. 
The resultant spatial distribution of the polarization  is controlled by the superposition of the eigenmodes:
%
%
\begin{equation} \label{eqn:LGLR}
E_1(\boldsymbol r,\boldsymbol\phi) = \epsilon_L LG_{p_L}^{\ell_L}  \, ; \, E_2(\boldsymbol r,\boldsymbol\phi) = \epsilon_R LG_{p_R}^{\ell_R} 
\end{equation}
where $\epsilon_L$ and $\epsilon_R$ are the field mode amplitudes, $p$ is the radial index, and $\ell$ is the orbital angular momentum (OAM) index of the mode~\cite{Yao:11}. Taking the modes at the beam waist $w_0$ and assuming $p = 0$, the $LG$ modes may be written as~\cite{Barnett07}:
\begin{equation}
LG_0^{\ell} (\boldsymbol r,\boldsymbol\phi)= \sqrt{\frac{2}{\pi w_0^2 |\ell | ! }}\left( \frac{\sqrt{2}r}{w_0} \right)^{| \ell |}\exp \left(-\frac{r^2}{w_0^2}+ i \ell \phi \right),
\label{eqn:LGmode}
\end{equation}
where: $r=\sqrt{x^2+y^2}$ is the radial coordinate, and $\phi=\tan^{-1}(y/x)$ is the azimuthal coordinate.

If either $E_1$ or $E_2$ is zero, then the resultant beam is an $LG$ mode with spatially uniform circular polarization. If the two modes have equal amplitudes and the same OAM ($\ell_L=\ell_R$), the resultant beam will have spatially uniform linear polarization. If they have equal amplitudes and equal but opposite OAM ($\ell_L=-\ell_R$), however, the resultant Cylindrical Vector (CV) beam~\cite{Zhan:09, Galvez:12} will have an azimuthally varying linear polarization distribution that may be radial, azimuthal or spiral, depending on the phase difference $\beta$. 
If the two modes have different magnitudes of OAM the resultant `full Poincar\'e ' beam will carry a net OAM and the polarization will vary in both the azimuthal and radial coordinates and may contain all states of polarization: linear; elliptical; and circular~\cite{Beckley:10}. Typical examples are the so-called `lemon' and `star' beams~\cite{Nye83}.
Note that for beams with Cartesian symmetry, the profiles may be better expressed in as Hermite-Gaussian ($HG$) modes \cite{Beijersbergen:93} and with linear polarization vectors $\boldsymbol{\hat{e}}_1=\boldsymbol{\hat{e}}_x$, $\boldsymbol{\hat{e}}_2=\boldsymbol{\hat{e}}_y$.

The generation of such beams commonly relies upon methods that use external conversion optics to superimpose orthogonally-polarised transverse modes, including interferometric techniques~\cite{Niziev:06,Chen:14}, q-plates~\cite{PhysRevLett.117.233903, Cardano:13}, and liquid crystal spatial light modulators~\cite{Maurer_2007}. 
CV beams have also been demonstrated in the ultraviolet using higher harmonic generation~\cite{Hernandez-Garcia:17}. 
\section{Generation of Poincar\'e beams using a free electron laser}
In this paper, a new FEL method for generating bright, tunable, coherent Poincar\'e beams is proposed without the need for any external conversion optics. This optics-free method allows the extension of Poincar\'e beam generation into the X-ray regime for the first time.  
It is shown that 
by combining techniques of polarization and transverse mode shaping with FEL `afterburners', coherent harmonic emission processes can be used to generate several classes of Poincar\'e beams - including radially polarized CV beams and `star' Poincar\'e beams - with minimal changes to the overall facility layout. This approach enables the generation of wavelength-tunable, narrowband X-ray FSL beams in modern FEL facilities providing, for example, high resolution spectroscopy or scanning over narrow atomic/molecular resonances with structured light pulses.


FELs use highly relativistic electron beams (e-beams) propagating through undulating magnetic fields (undulators) to generate intense, tunable pulses of light. The wavelength range of FELs is broad and easily tunable, with current shorter wavelength facilities operating from the VUV down to hard X-rays~\cite{ackermann2007operation,emma2010first,ishikawa2012compact,allaria2012highly,FERMI2stage,PALFEL}. 
The radiation output is typically a transverse Gaussian mode with nearly full transverse coherence and a spatially homogeneous polarization that is determined by the magnetic undulator fields (planar, helical, or elliptical).
Polarization control 
is thus enabled by undulators with tunable polarity~(e.g. \cite{PhysRevX.4.041040,photonics4020029,ellipticNJP}), or by a short tunable undulator section placed immediately downstream. This `afterburner' undulator uses an FEL e-beam that has a strong coherent bunching from the previous lasing stage to generate coherent light with a high degree of adjustable polarization~\cite{lutman2016polarization}. The primary FEL radiation pulse energy can also be strongly suppressed (but the e-beam bunching preserved) by using an undulator with reverse-tapering~\cite{PhysRevSTAB.16.110702} and by e-beam steering~\cite{MacArthur2018PhysRevX}, so that only the radiation pulse generated in the afterburner is delivered to experiments.


Currently, such schemes rely on transverse Gaussian afterburner emission at the first harmonic in which the e-beam bunching wavelength, $\lambda_b$, matches the fundamental afterburner radiation wavelength resonance, $\lambda_b=\lambda_r=\lambda_u(1+K^2)/2\gamma^2$, where $\lambda_u$ is the afterburner period, $K$ is its rms undulator parameter, and $\gamma$ is the e-beam relativistic factor. By radiating at  harmonics however, where $\lambda_b=\lambda_r/h$ with $h>1$ an integer, the transverse mode properties of the afterburner emission can be tailored to enable the generation of FSL beams. For example, in helical undulators that generate circularly polarized light, the coherent emission at harmonics is well-characterized by an $LG$ mode with a helical phase and OAM index $\ell=\mp(h-1)$~\cite{sasaki2008proposal}. Both the sign of $\ell$ and the circular polarization vector $(\boldsymbol{\hat{e}}_x\mp i\boldsymbol{\hat{e}}_y)/\sqrt{2}$ of the radiation are determined by the direction of the e-beam trajectory and therefore on the left (-) or right (+) handedness of the undulator. For planar undulators, the emission is linearly polarised, and the harmonic intensity profiles resemble an $HG$ mode basis set. 

 \begin{figure}{\mathindent=0pt}
    \centering
\includegraphics[width=0.6\textwidth]{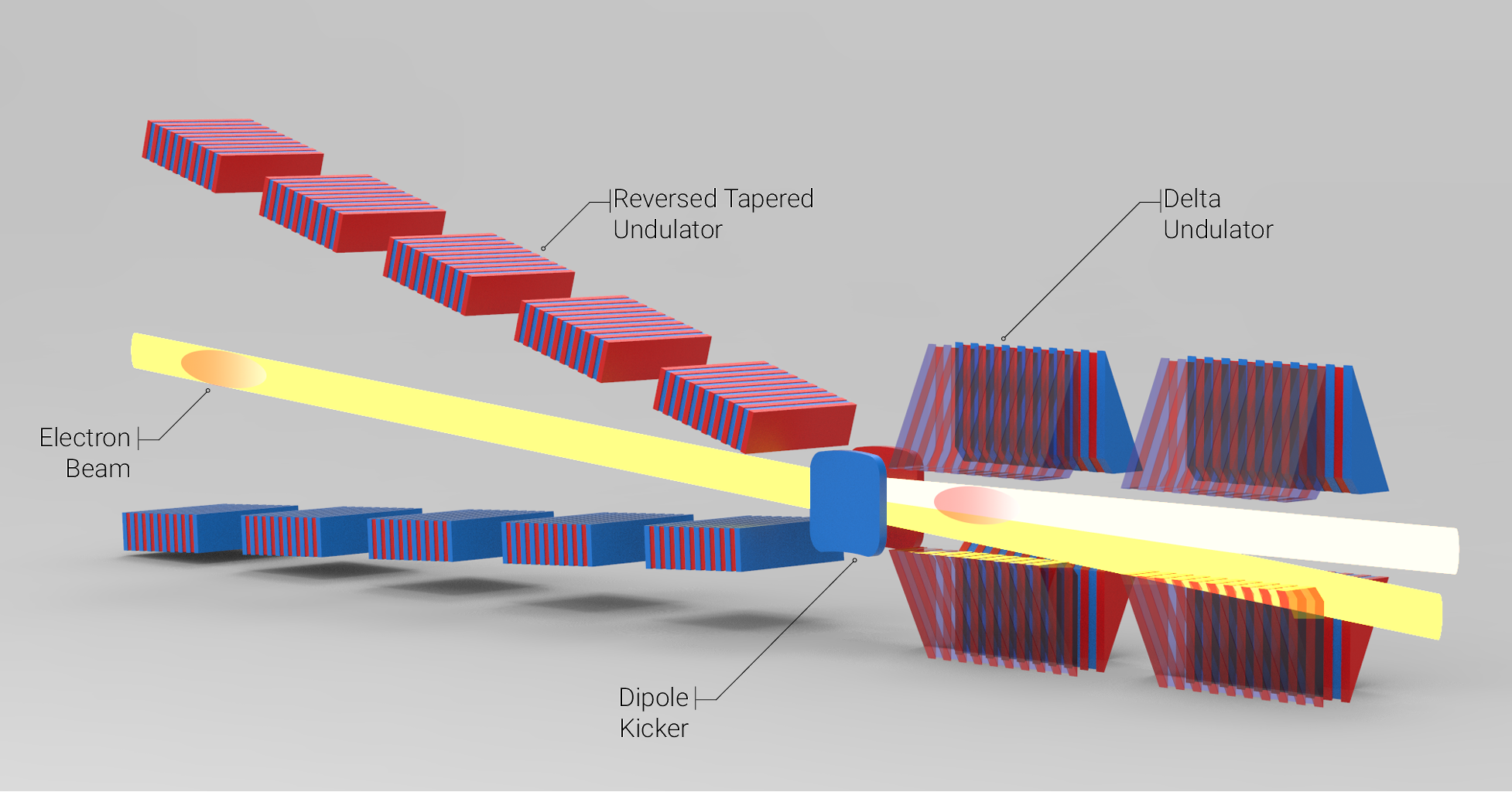}
 \caption[width=\textwidth]{Schematic of the method. A relativistic electron beam is initially bunched using a reverse-tapered undulator. This suppresses the generation of the linearly polarised radiation. A dipole kicker (or off-axis quadrupole lattice) then redirects the electron beam into two Delta undulators tuned so that the electron bunching is at an harmonic of their fundamental resonant wavelength. The Delta undulators can then be adjusted for different polarizations and tuning to generate light with transversely inhomogeneous polarization - Fully Structured Light.}
\label{fig:setup}
\end{figure}


Here, we propose extending this concept to two consecutive orthogonally polarised afterburners, individually adjustable in their strength, polarization, and relative phase. The two radiation pulses emitted from each of the two undulators overlap spatially and temporally. Previous versions of this crossed undulator setup have experimentally demonstrated polarization control at the fundamental~\cite{ferrari2019free, deng2014polarization}. In the method described here, the electrons emit higher-order transverse modes at the second harmonic in one or both of the afterburners. This results in an FSL beam with an output field described by the superpostion of modes, as in Eq.~\ref{eq:vector superposition of transverse modes}. The phase difference, $\beta$, between the two modes can also be controlled by using a small magnetic phase-shifter between the two afterburners. This also allows phase-shifts due to the slippage between modules to be compensated, or the polarization pattern to be modified or rotated. Such a setup can be constructed from two Delta-type (or Apple-II) afterburner undulators~\cite{SASAKI199483,PhysRevSTAB.11.120702}, provided that they have sufficient adjustment of their $K$ parameters to access harmonics - extending the practicality of these undulators to enable transverse  polarization control.

\section{Simulations Results}
The FEL simulation code Puffin~\cite{campbell2012puffin} is used to model the setup shown in Fig.~\ref{fig:setup}. In contrast to most other FEL codes, Puffin does not average the electron motion over an undulator period, allowing modeling of both planar and helical harmonic emission arising from electron motion at the sub-undulator period scale. The setup is modelled using parameters based on the LCLS-II project at SLAC~\cite{LCLSIIScience}, with electron beam energy $4$~GeV, peak current $I_0=1$~kA, and beam radius $\sigma_x=26~\mu$m. The undulator period is $\lambda_u =3.9$~cm and each afterburner section has $N_u$=20 undulator periods.

Time-independent (steady-state) simulations were used to demonstrate the method. This mode does not model the full temporal duration of the electron beam. However, as demonstrated in other crossed undulator methods, temporal pulse effects should not significantly affect the results, as the bunching factor on entering the afterburners is orders of magnitude larger than any beam shot-noise, and the relative slippage between electrons and radiation pulses is less than the coherence length~\cite{li20103d}. The electrons are first pre-bunched in a reverse-tapered FEL section with $\lambda_b=1.25$~nm. This achieves a bunching factor $|b|=0.45$, while also reducing the FEL output power to 1~MW, three orders of magnitude lower than without the undulator taper. The process for pre-bunching electrons in a reverse tapered undulator has been described previously \cite{PhysRevSTAB.16.110702}. The electron beam bunching process does not differ significantly from the standard FEL process with the exception of the reduced radiation power. Steering the e-beam to further reduce the contribution from the background power is not modeled~\cite{MacArthur2018PhysRevX}, and the radiation generated in the FEL section is simply removed. The pre-bunched beam then enters the afterburner Delta undulators, which can be adjusted for linear or circular polarization and tuned so that the electron bunching is at either the fundamental or second harmonic. Three polarization distributions that generate FSL beams are now presented using this setup.

\subsection{Vector beams}
\begin{figure}
\centering
\includegraphics[width=0.6\textwidth]{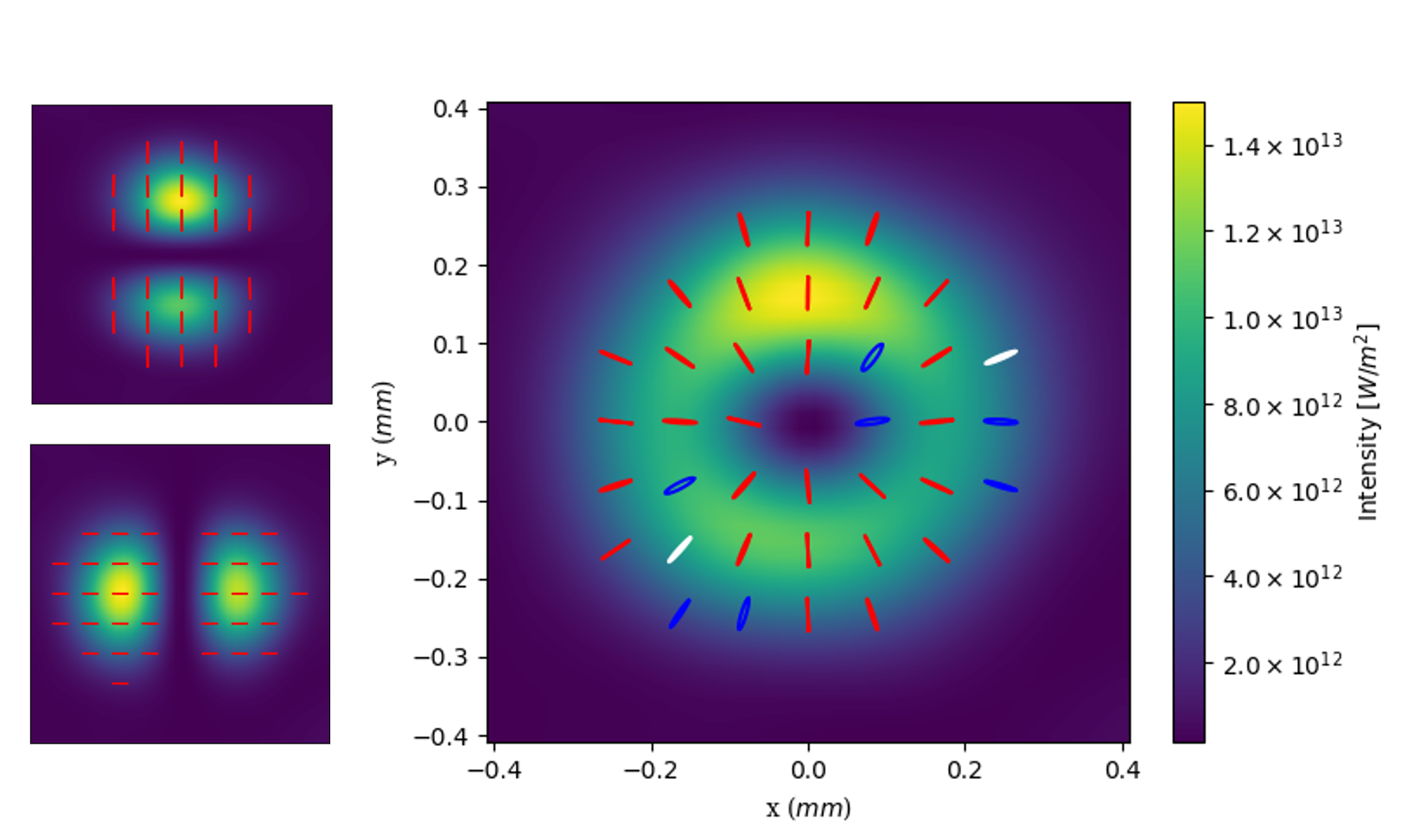}
\includegraphics[width=0.6\textwidth]{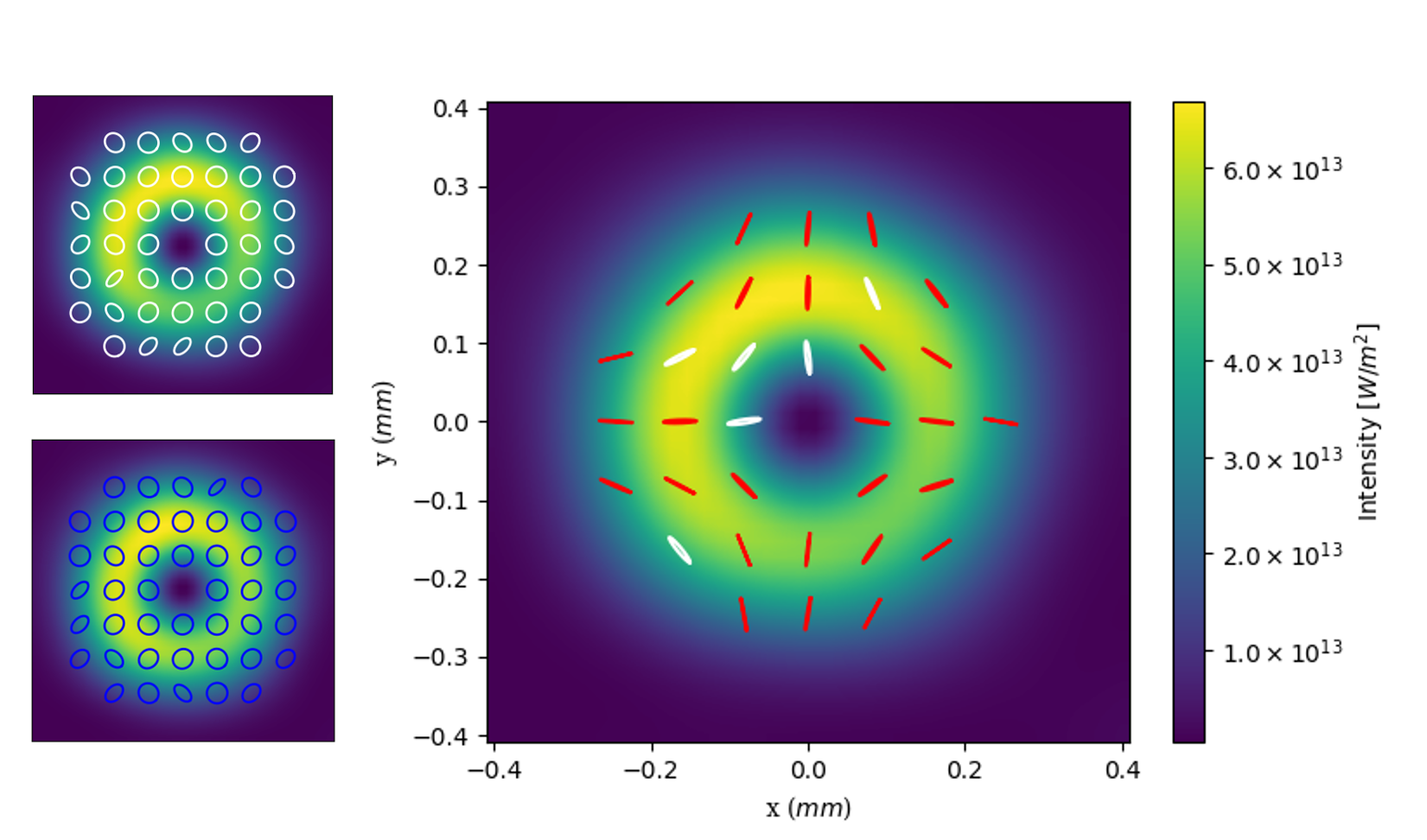}
\caption{\label{fig:Vector Beams} Simulation of cylindrical vector polarizations. The electrons are bunched at the second harmonic of the Delta undulators to give  orthogonal linear (top) and circular (bottom) polarization. The radiation polarization emitted from each Delta undulator is plotted schematically on the (left) two plots and the combined field simulated from both undulators on the (right) plot. Red, blue and white lines correspond to linear, right-circular and left-circular polarization respectively.}
\end{figure}

In the first example, a pair of cross-polarized planar afterburners is simulated. They are both tuned to a fundamental resonance of $\lambda_r=2.5$~nm, so that the e-beam is bunched and radiates at the second harmonic, generating the field 
$\boldsymbol{E}(\boldsymbol r,\boldsymbol\phi)=\epsilon_1 HG_{10}\boldsymbol{\hat{e}_x}+e^{i\beta}\epsilon_2 HG_{01}\boldsymbol{\hat{e}_y}$. 
With $\epsilon_1=\epsilon_2$, this superposition is seen in Fig.~\ref{fig:Vector Beams} to create an annular intensity profile with a radial polarization distribution for $\beta=0$. The polarization map was constructed by calculating the Stokes parameters:
\begin{align}
S_0 &= |E_x|^2+|E_y|^2 = |E_R|^2+|E_L|^2 , \nonumber\\  
S_1 &= |E_x|^2-|E_y|^2 = 2 \operatorname{Re}(E_R^*E_L) , \nonumber\\
S_2 &= 2\operatorname{Re}(E_x^*E_y) = 2\operatorname{Im}(E_R^*E_L) , \nonumber\\      
S_3 &= 2\operatorname{Im}(E_x^*E_y) = |E_R|^2-|E_L|^2 ,
\end{align}

where the subscripts denote the appropriate linear or circular field components~\cite{Dennis:09}. $S_0$ is the parameter describing temporal intensity. The linear horizontal/vertical, diagonal linear, and circular polarization are described by $S_1$, $S_2$, and $S_3$, respectively.

The ellipticity, $\chi$, and the orientation, $\psi$, of a polarization ellipse at each point on the transverse plane can then be calculated~\cite{Saleh}, where: 
\begin{equation}
\chi=\frac{1}{2}\sin^{-1}{\left(\frac{S_3}{S_0}\right)}  \, , \,
\psi=\frac{1}{2}\tan^{-1}{\left(\frac{S_2}{S_1}\right)} .
\end{equation}

The polarization ellipses are then plotted at various points across the intensity profile. The normalised stokes vector capturing the spatial polarization for these crossed planar harmonic undulators can be written as, 
\begin{equation}\label{eq:helical undulator superposition}
    \boldsymbol{s}=\frac{1}{S_0}\begin{pmatrix}S_1\\S_2\\S_3\end{pmatrix}=\begin{pmatrix} 
\cos(2\phi) \\
\cos(\beta)\sin(2\phi)\\
\sin(\beta)\sin(2\phi)\\
\end{pmatrix}.
\end{equation}
Note that for $\beta\neq0$, the polarization distribution can also contain circular components.

Similarly, with crossed helical undulators tuned so that the bunching is at the second harmonic, one obtains ${\ell_R=-1}$ and $\ell_L=1$. If the undulators are the same length and the bunching factor does not change significantly between them, the modes have equal amplitude, $\epsilon_L=\epsilon_R$. We then obtain, 
\begin{equation}\label{eq:helical undulator superposition2}
    \boldsymbol{s}=\begin{pmatrix}\cos(\beta-2\phi)\\
    \sin(\beta-2\phi)\\
    0\end{pmatrix}.
\end{equation}
The $S_3$ parameter vanishes, so the beam has only linear polarization states which vary with $\phi$. The generated vector `vortex beam' is also shown in Fig.~\ref{fig:Vector Beams}.

We note that, in order for this description to accurately model the final FSL output, the radiation emitted in each undulator should be well-described by a pure modes. 
In Ref.~\cite{PhysRevAccelBeams.23.020703}, it was shown that with sufficiently large $K$ and periods $N_u$ in a helical afterburner, coherent radiation from a pre-bunched e-beam is well approximated by an $LG$ mode in the limit that the e-beam radius satisfies $\sigma_x > \gamma_z\sqrt{N_u}/k$, such that the emission angles are dominated by the e-beam and not the undulator emission. 
The undulators must also be kept relatively short to reduce the diffraction of the radiation so that the the transverse sizes of the modes do not significantly differ.

It is  seen from Fig.~\ref{fig:Vector Beams} that CV beams are generated when the orthogonal afterburners both radiate at the second harmonic. Due to the relationship between the polarization and the transverse modes, only certain CV polarization distributions are available using this setup. For example, the second harmonic emission does not produce $y$-polarised $HG_{10}$ modes, or $\ell_L=-1$ modes with left-circular polarization (i.e., no `lemon' beams)~\cite{afanasev2011generation,Katoh2017prl}.

The power of the final radiation pulses in Fig.~\ref{fig:Vector Beams} is of the order of 0.3~MW, which is consistent with the second harmonic power calculated in~\cite{PhysRevAccelBeams.23.020703}. For a single helical afterburner the radiated coherent power is
\begin{equation}
    P=4P_b b^2\frac{I_0}{\gamma I_A}\left(\frac{K^2}{1+K^2}\right)^2\ln\left(\frac{1+4N^2}{4N^2}\right) 
\label{pow}
\end{equation}
where $P_b$ is the peak e-beam power, $I_A$= 17 kA is the Alfven current, and $N=k\sigma_x^2/L_u$ is the Fresnel number of the e-beam with $k=2\pi/\lambda_b$, and $L_u=N_u\lambda_u$ the length of the undulator. This power is the same magnitude as the radiation emitted at the fundamental of the upstream reverse-tapered FEL, highlighting the need to steer the pre-bunched electron beam to avoid overlap with the radiation emitted during pre-bunching. We note from Eqn.~(\ref{pow}) that with strong focussing to reduce $\sigma_x<26~\mu$m, the power output can be greatly improved. As the electrons are bunched before the afterburner, the short undulators needed to account for diffraction still provide high powers, although the power scaling for a single undulator favors small $N$. If a longer afterburner section is desirable, or if required parameters lead to greater diffraction, one solution is to split the first Delta undulator into two sections and sandwich the second Delta undulator between these two sections. This leads to better overlap of the two polarised beams. 

\subsection{Full Poincar\'e beams}

\begin{figure}
\centering
\includegraphics[width=0.6\textwidth]{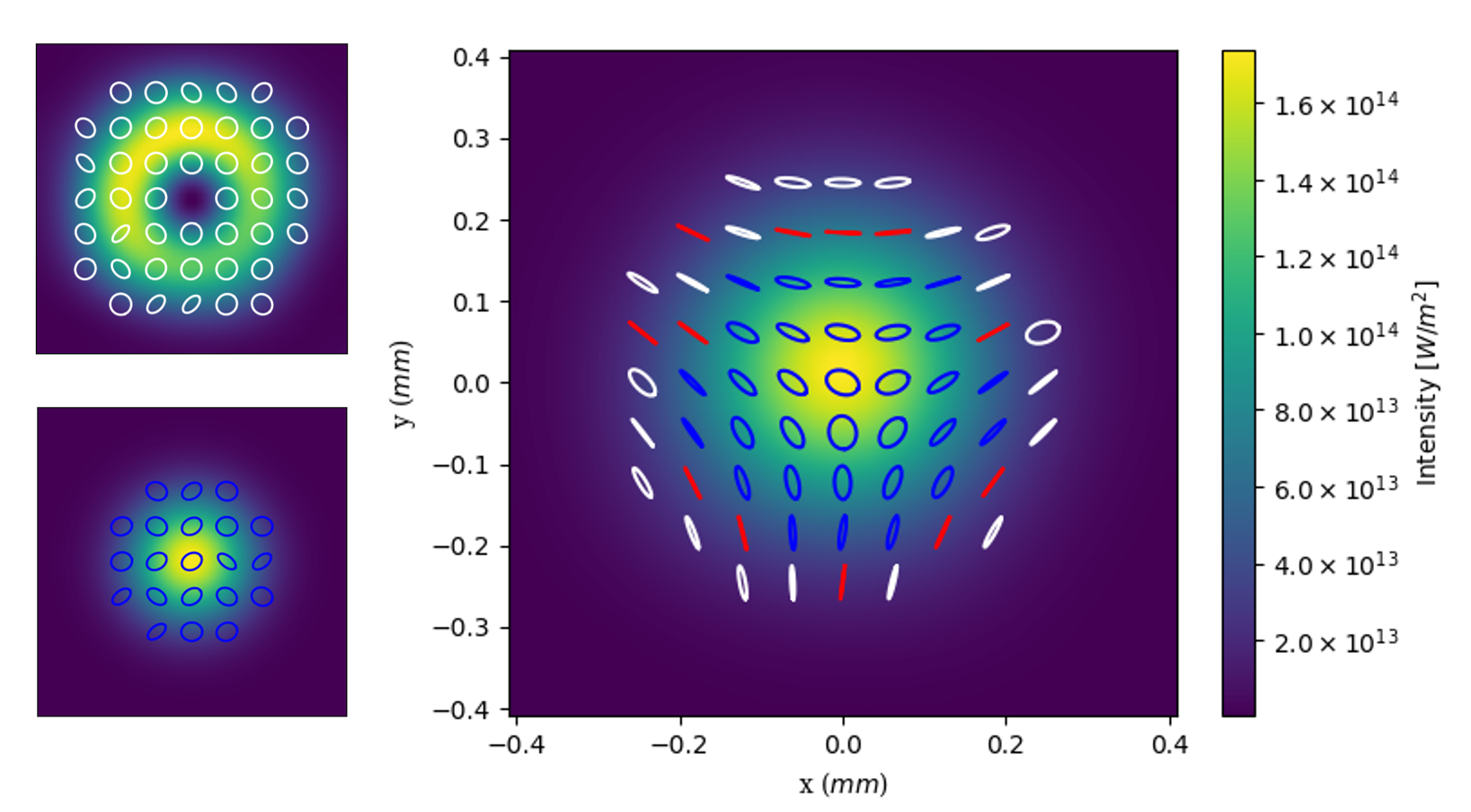}
 \caption{\label{fig:Poinecare_Beam} Poincar\'e polarization distribution downstream of the afterburner. The Delta undulators are set to have orthogonal circular polarizations. The electrons are bunched at the second harmonic of the first undulator and the fundamental of the second undulator. The radiation emitted in each Delta undulator is plotted (left) along with the combined field from both undulators  (right).}
 \end{figure}

The second class of light with spatially inhomogeneous polarization considered is full Poincar\'e beams created from a superposition of $LG_0^{\pm1}$ and $LG_0^{0}$ (Gaussian) radiation with orthogonal circular polarizations. 
From Eqn.~(\ref{eqn:LGLR}), the Stokes vector then becomes, 
\begin{equation}
    \boldsymbol{s}=\begin{pmatrix}\frac{2\sqrt{2}rw_0}{2r^2+w_0^2}\cos(\beta-\phi)\\\frac{2\sqrt{2}rw_0}{2r^2+w_0^2}\sin(\beta-\phi)\\\pm\frac{2r^2-w_0^2}{2r^2+w_0^2}\end{pmatrix},
\end{equation}
where the $+$ and $-$ signs correspond to $(\ell_L,\ell_R)=(1,0)$ and $(\ell_L,\ell_R)=(0,-1)$, respectively. On axis, $r=0$, the polarization is purely circular while the at the radius $r=w_0/\sqrt{2}$, the polarization is purely linear, with orientation depending on $\phi$. Fig.~\ref{fig:Poinecare_Beam} shows the `star' Poincar\'e beam output generated in the $(\ell_L,\ell_R)=(1,0)$ case. 
To achieve this combination, the second undulator is tuned so that its fundamental resonance matches the bunching wavelength at $\lambda_r=1.25$~nm and the radiation emitted is Gaussian. The first undulator is tuned to $\lambda_r=2.5$~nm, radiating at the second harmonic as before. The electrons radiate with higher power at the fundamental than at the second harmonic. To compensate and balance the powers between the two radiation beams, the Delta undulator emitting at the fundamental is detuned from resonance to reduce its power output. Detuning the undulator will affect the mode size and therefore polarization structure. Specific undulator detuning is a topic for future studies and will depend on the specific application.

\section{Conclusion}

The three x-ray polarization topologies demonstrated here are not the full range of pulses available with the two Delta undulator arrangement. In addition to varying the polarization and undulator resonance, other factors can change the polarization distribution. Both the phase and power ratio between the different transverse modes can be adjusted which, for example, could be used to create elliptical vector beams. Slightly detuning the resonance of one undulator will push the radiation further off axis, which can be used to control the mode overlap~\cite{ferrari2019free}. Finally, radiating at even higher harmonics of a helical undulator will generate the higher order $LG$ modes producing yet more variants, though the power drops with increasing harmonic number~\cite{PhysRevAccelBeams.23.020703}.

We note that this method can generate Poincar\'e beams at any operational wavelength of a FEL facility. The advantage of the afterburner configuration is that it is both simple and cost effective to implement as the afterburners can be added to existing FEL facilities, or may already exist as the last couple undulator sections. Furthermore, the method could be combined with other methods. For example, consideration of temporal or short pulse effects can be envisaged that alter the FSL in the temporal domain (e.g., \cite{SerkezPRABpolshaping2019}) adding a further dimension for potential experimental exploitation.  
\section*{Acknowledgments}
We are grateful to funding from the Science and Technology Facilities Council (Agreement Number 4163192 Release\#3); ARCHIEWeSt HPC, EPSRC grant EP/K000586/1; EPSRC Grant EP/M011607/1; John von Neumann Institute for Computing (NIC) on JUROPA at J\"ulich Supercomputing Centre (JSC), project HHH20. E.H. was supported by U.S. Department of Energy Contract No. DE-AC02-76SF00515 and award no. 2017-SLAC-100382.\\

\section*{References}


\begin{thebibliography}{10}
\expandafter\ifx\csname url\endcsname\relax
  \def\url#1{{\tt #1}}\fi
\expandafter\ifx\csname urlprefix\endcsname\relax\def\urlprefix{URL }\fi
\providecommand{\eprint}[2][]{\url{#2}}

\bibitem{Zhan:09}
Zhan Q 2009 {\em Adv. Opt. Photon.\/} {\bf 1} 1--57
  \urlprefix\url{http://aop.osa.org/abstract.cfm?URI=aop-1-1-1}

\bibitem{Beckley:10}
Beckley A~M, Brown T~G and Alonso M~A 2010 {\em Opt. Express\/} {\bf 18}
  10777--10785
  \urlprefix\url{http://www.opticsexpress.org/abstract.cfm?URI=oe-18-10-10777}

\bibitem{Galvez:12}
Galvez E~J, Khadka S, Schubert W~H and Nomoto S 2012 {\em Appl. Opt.\/} {\bf
  51} 2925--2934
  \urlprefix\url{http://ao.osa.org/abstract.cfm?URI=ao-51-15-2925}

\bibitem{PhysRevLett.91.233901}
Dorn R, Quabis S and Leuchs G 2003 {\em Phys. Rev. Lett.\/} {\bf 91}(23) 233901
  \urlprefix\url{https://link.aps.org/doi/10.1103/PhysRevLett.91.233901}

\bibitem{Rubinsztein_Dunlop_2016}
Rubinsztein-Dunlop H, Forbes A, Berry M~V, Dennis M~R, Andrews D~L, Mansuripur
  M, Denz C, Alpmann C, Banzer P, Bauer T, Karimi E, Marrucci L, Padgett M,
  Ritsch-Marte M, Litchinitser N~M, Bigelow N~P, Rosales-Guzm{\'{a}}n C,
  Belmonte A, Torres J~P, Neely T~W, Baker M, Gordon R, Stilgoe A~B, Romero J,
  White A~G, Fickler R, Willner A~E, Xie G, McMorran B and Weiner A~M 2016 {\em
  Journal of Optics\/} {\bf 19} 013001

\bibitem{PhysRevLett.117.233903}
Bouchard F, Larocque H, Yao A~M, Travis C, De~Leon I, Rubano A, Karimi E, Oppo
  G~L and Boyd R~W 2016 {\em Phys. Rev. Lett.\/} {\bf 117}(23) 233903
  \urlprefix\url{https://link.aps.org/doi/10.1103/PhysRevLett.117.233903}

\bibitem{Nesterov_2000}
Nesterov A~V and Niziev V~G 2000 {\em Journal of Physics D: Applied Physics\/}
  {\bf 33} 1817--1822
  \urlprefix\url{https://doi.org/10.1088%2F0022-3727%2F33%2F15%2F310}

\bibitem{Hao_2010}
Hao X, Kuang C, Wang T and Liu X 2010 {\em Journal of Optics\/} {\bf 12} 115707
  \urlprefix\url{https://doi.org/10.1088%2F2040-8978%2F12%2F11%2F115707}

\bibitem{PhysRevLett.85.4482}
Sick B, Hecht B and Novotny L 2000 {\em Phys. Rev. Lett.\/} {\bf 85}(21)
  4482--4485
  \urlprefix\url{https://link.aps.org/doi/10.1103/PhysRevLett.85.4482}

\bibitem{PhysRevLett.86.5251}
Novotny L, Beversluis M~R, Youngworth K~S and Brown T~G 2001 {\em Phys. Rev.
  Lett.\/} {\bf 86}(23) 5251--5254
  \urlprefix\url{https://link.aps.org/doi/10.1103/PhysRevLett.86.5251}

\bibitem{McNeilNatPho2010}
McNeil B~W~J and Thompson N~R 2010 {\em Nature Photonics\/} {\bf 4} 814--821
  \urlprefix\url{https://doi.org/10.1038/nphoton.2010.239}

\bibitem{chiralReview}
Cameron R~P, G\"otte J~B, Barnett S~M and Yao A~M 2017 {\em Phil. Trans. R.
  Soc. A\/} {\bf 375} 20150433

\bibitem{Yao:11}
Yao A~M and Padgett M~J 2011 {\em Adv. Opt. Photon.\/} {\bf 3} 161--204
  \urlprefix\url{http://aop.osa.org/abstract.cfm?URI=aop-3-2-161}

\bibitem{Barnett07}
Barnett S~M and Zambrini R 2007 {\em Quantum Imaging, Ed. K. I. Kobolov\/}
  (Springer, Singapore)

\bibitem{Nye83}
Nye J~F 1983 {\em Proc. R Soc. Lond. A\/} {\bf 389} 279

\bibitem{Beijersbergen:93}
Beijersbergen M, Allen L, van~der Veen H and Woerdman J 1993 {\em Optics
  Communications\/} {\bf 96} 123 -- 132 ISSN 0030-4018
  \urlprefix\url{http://www.sciencedirect.com/science/article/pii/003040189390535D}

\bibitem{Niziev:06}
Niziev V~G, Chang R~S and Nesterov A~V 2006 {\em Appl. Opt.\/} {\bf 45}
  8393--8399 \urlprefix\url{http://ao.osa.org/abstract.cfm?URI=ao-45-33-8393}

\bibitem{Chen:14}
Chen S, Zhou X, Liu Y, Ling X, Luo H and Wen S 2014 {\em Opt. Lett.\/} {\bf 39}
  5274--5276 \urlprefix\url{http://ol.osa.org/abstract.cfm?URI=ol-39-18-5274}

\bibitem{Cardano:13}
Cardano F, Karimi E, Marrucci L, de~Lisio C and Santamato E 2013 {\em Opt.
  Express\/} {\bf 21} 8815--8820
  \urlprefix\url{http://www.opticsexpress.org/abstract.cfm?URI=oe-21-7-8815}

\bibitem{Maurer_2007}
Maurer C, Jesacher A, F\"{u}rhapter S, Bernet S and Ritsch-Marte M 2007 {\em
  New Journal of Physics\/} {\bf 9} 78--78

\bibitem{Hernandez-Garcia:17}
Hern\'{a}ndez-Garc\'{i}a C, Turpin A, Rom\'{a}n J~S, Pic\'{o}n A, Drevinskas R,
  Cerkauskaite A, Kazansky P~G, Durfee C~G and {n}igo J~Sola I 2017 {\em
  Optica\/} {\bf 4} 520--526
  \urlprefix\url{http://www.osapublishing.org/optica/abstract.cfm?URI=optica-4-5-520}

\bibitem{ackermann2007operation}
Ackermann W, Asova G, Ayvazyan V, Azima A, Baboi N, B{\"a}hr J, Balandin V,
  Beutner B, Brandt A, Bolzmann A {\em et~al.\/} 2007 {\em Nature photonics\/}
  {\bf 1} 336--342

\bibitem{emma2010first}
Emma P, Akre R, Arthur J, Bionta R, Bostedt C, Bozek J, Brachmann A, Bucksbaum
  P, Coffee R, Decker F~J {\em et~al.\/} 2010 {\em nature photonics\/} {\bf 4}
  641

\bibitem{ishikawa2012compact}
Ishikawa T, Aoyagi H, Asaka T, Asano Y, Azumi N, Bizen T, Ego H, Fukami K,
  Fukui T, Furukawa Y {\em et~al.\/} 2012 {\em nature photonics\/} {\bf 6}
  540--544

\bibitem{allaria2012highly}
Allaria E, Appio R, Badano L, Barletta W, Bassanese S, Biedron S, Borga A,
  Busetto E, Castronovo D, Cinquegrana P {\em et~al.\/} 2012 {\em Nature
  Photonics\/} {\bf 6} 699--704

\bibitem{FERMI2stage}
Allaria E, Castronovo D, Cinquegrana P, Craievich P, Dal~Forno M, Danailov M~B,
  D'Auria G, Demidovich A, De~Ninno G, Di~Mitri S, Diviacco B, Fawley W~M,
  Ferianis M, Ferrari E, Froehlich L, Gaio G, Gauthier D, Giannessi L, Ivanov
  R, Mahieu B, Mahne N, Nikolov I, Parmigiani F, Penco G, Raimondi L, Scafuri
  C, Serpico C, Sigalotti P, Spampinati S, Spezzani C, Svandrlik M, Svetina C,
  Trovo M, Veronese M, Zangrando D and Zangrando M 2013 {\em Nature
  Photonics\/} {\bf 7} 913 EP --
  \urlprefix\url{https://doi.org/10.1038/nphoton.2013.277}

\bibitem{PALFEL}
Kang H~S, Min C~K, Heo H, Kim C, Yang H, Kim G, Nam I, Baek S~Y, Choi H~J, Mun
  G, Park B~R, Suh Y~J, Shin D~C, Hu J, Hong J, Jung S, Kim S~H, Kim K, Na D,
  Park S~S, Park Y~J, Han J~H, Jung Y~G, Jeong S~H, Lee H~G, Lee S, Lee S, Lee
  W~W, Oh B, Suh H~S, Parc Y~W, Park S~J, Kim M~H, Jung N~S, Kim Y~C, Lee M~S,
  Lee B~H, Sung C~W, Mok I~S, Yang J~M, Lee C~S, Shin H, Kim J~H, Kim Y, Lee
  J~H, Park S~Y, Kim J, Park J, Eom I, Rah S, Kim S, Nam K~H, Park J, Park J,
  Kim S, Kwon S, Park S~H, Kim K~S, Hyun H, Kim S~N, Kim S, Hwang S~m, Kim M~J,
  Lim C~y, Yu C~J, Kim B~S, Kang T~H, Kim K~W, Kim S~H, Lee H~S, Lee H~S, Park
  K~H, Koo T~Y, Kim D~E and Ko I~S 2017 {\em Nature Photonics\/} {\bf 11}
  708--713 \urlprefix\url{https://doi.org/10.1038/s41566-017-0029-8}

\bibitem{PhysRevX.4.041040}
Allaria E, Diviacco B, Callegari C, Finetti P, Mahieu B, Viefhaus J, Zangrando
  M, De~Ninno G, Lambert G, Ferrari E, Buck J, Ilchen M, Vodungbo B, Mahne N,
  Svetina C, Spezzani C, Di~Mitri S, Penco G, Trov\'o M, Fawley W~M, Rebernik
  P~R, Gauthier D, Grazioli C, Coreno M, Ressel B, Kivim\"aki A, Mazza T,
  Glaser L, Scholz F, Seltmann J, Gessler P, Gr\"unert J, De~Fanis A, Meyer M,
  Knie A, Moeller S~P, Raimondi L, Capotondi F, Pedersoli E, Plekan O, Danailov
  M~B, Demidovich A, Nikolov I, Abrami A, Gautier J, L\"uning J, Zeitoun P and
  Giannessi L 2014 {\em Phys. Rev. X\/} {\bf 4}(4) 041040
  \urlprefix\url{https://link.aps.org/doi/10.1103/PhysRevX.4.041040}

\bibitem{photonics4020029}
Roussel E, Allaria E, Callegari C, Coreno M, Cucini R, Mitri S~D, Diviacco B,
  Ferrari E, Finetti P, Gauthier D, Penco G, Raimondi L, Svetina C, Zangrando
  M, Beckmann A, Glaser L, Hartmann G, Scholz F, Seltmann J, Shevchuk I,
  Viefhaus J and Giannessi L 2017 {\em Photonics\/} {\bf 4} ISSN 2304-6732
  \urlprefix\url{https://www.mdpi.com/2304-6732/4/2/29}

\bibitem{ellipticNJP}
Henderson J, Campbell L, Freund H and McNeil B 2016 {\em New Journal of
  Physics\/} {\bf 18} 062003

\bibitem{lutman2016polarization}
Lutman A~A, MacArthur J~P, Ilchen M, Lindahl A~O, Buck J, Coffee R~N, Dakovski
  G~L, Dammann L, Ding Y, D{\"u}rr H~A {\em et~al.\/} 16 {\em Nature
  photonics\/} {\bf 10} 468

\bibitem{PhysRevSTAB.16.110702}
Schneidmiller E~A and Yurkov M~V 2013 {\em Phys. Rev. ST Accel. Beams\/} {\bf
  16}(11) 110702
  \urlprefix\url{https://link.aps.org/doi/10.1103/PhysRevSTAB.16.110702}

\bibitem{MacArthur2018PhysRevX}
MacArthur J~P, Lutman A~A, Krzywinski J and Huang Z 2018 {\em Phys. Rev. X\/}
  {\bf 8}(4) 041036
  \urlprefix\url{https://link.aps.org/doi/10.1103/PhysRevX.8.041036}

\bibitem{sasaki2008proposal}
Sasaki S and McNulty I 2008 {\em Physical review letters\/} {\bf 100} 124801

\bibitem{ferrari2019free}
Ferrari E, Roussel E, Buck J, Callegari C, Cucini R, De~Ninno G, Diviacco B,
  Gauthier D, Giannessi L, Glaser L {\em et~al.\/} 2019 {\em Physical Review
  Accelerators and Beams\/} {\bf 22} 080701

\bibitem{deng2014polarization}
Deng H, Zhang T, Feng L, Feng C, Liu B, Wang X, Lan T, Wang G, Zhang W, Liu X
  {\em et~al.\/} 2014 {\em Physical Review Special Topics-Accelerators and
  Beams\/} {\bf 17} 020704

\bibitem{SASAKI199483}
Sasaki S 1994 {\em Nuclear Instruments and Methods in Physics Research Section
  A: Accelerators, Spectrometers, Detectors and Associated Equipment\/} {\bf
  347} 83 -- 86 ISSN 0168-9002
  \urlprefix\url{http://www.sciencedirect.com/science/article/pii/0168900294918597}

\bibitem{PhysRevSTAB.11.120702}
Temnykh A~B 2008 {\em Phys. Rev. ST Accel. Beams\/} {\bf 11}(12) 120702
  \urlprefix\url{https://link.aps.org/doi/10.1103/PhysRevSTAB.11.120702}

\bibitem{campbell2012puffin}
Campbell L and McNeil B 2012 {\em Physics of Plasmas\/} {\bf 19} 093119

\bibitem{LCLSIIScience}
Schoenlein R~W 2015 {New Science Opportunities Enabled by LCLS-II X-ray Lasers}
  Tech. Rep. SLAC-R-1053 {SLAC National Accelerator Laboratory}

\bibitem{li20103d}
Li Y, Faatz B and Pfl\"ueger J 2010 {\em DESY print TESLA-FEL\/}

\bibitem{Dennis:09}
M~Dennis K~O and Padgett M 2009 {\em Prog. Opt.\/} {\bf 53} 293--363

\bibitem{Saleh}
Saleh B~E~A and Teich M~C 2007 {\em Fundamentals of Photonics\/} (Wiley, New
  York)

\bibitem{PhysRevAccelBeams.23.020703}
Hemsing E 2020 {\em Phys. Rev. Accel. Beams\/} {\bf 23}(2) 020703
  \urlprefix\url{https://link.aps.org/doi/10.1103/PhysRevAccelBeams.23.020703}

\bibitem{afanasev2011generation}
Afanasev A and Mikhailichenko A 2011 On generation of photons carrying orbital
  angular momentum in the helical undulator (\textit{Preprint}
  \eprint{1109.1603})

\bibitem{Katoh2017prl}
Katoh M, Fujimoto M, Kawaguchi H, Tsuchiya K, Ohmi K, Kaneyasu T, Taira Y,
  Hosaka M, Mochihashi A and Takashima Y 2017 {\em Phys. Rev. Lett.\/} {\bf
  118}(9) 094801
  \urlprefix\url{https://link.aps.org/doi/10.1103/PhysRevLett.118.094801}

\bibitem{SerkezPRABpolshaping2019}
Serkez S, Trebushinin A, Veremchuk M and Geloni G 2019 {\em Phys. Rev. Accel.
  Beams\/} {\bf 22}(11) 110705
  \urlprefix\url{https://link.aps.org/doi/10.1103/PhysRevAccelBeams.22.110705}

\end{thebibliography}
\providecommand{\noopsort}[1]{}\providecommand{\singleletter}[1]{#1}%
\providecommand{\newblock}{}

\end{document}